\documentclass[10pt,conference]{IEEEtran}
\IEEEoverridecommandlockouts
\usepackage[noadjust]{cite}
\usepackage{url}
\usepackage{xurl}
\usepackage[colorlinks=true,allcolors=blue]{hyperref}
\usepackage{enumitem}
\usepackage{tabularx}
\usepackage{amsmath,amsfonts}
\usepackage{graphicx}
\usepackage{booktabs}
\usepackage{textcomp}
\usepackage{xcolor}
\usepackage{adjustbox}
\usepackage{multirow}
\usepackage{longtable}
\usepackage{svg}
\usepackage{caption}
\usepackage{subcaption}
\usepackage{subfiles}
\usepackage{dirtytalk}
\usepackage[T1]{fontenc}
\usepackage{balance}
\begin{document}

\title{Contrastive Learning for API Aspect Analysis}

\author{\IEEEauthorblockN{G. M. Shahariar\IEEEauthorrefmark{1},
Tahmid Hasan\IEEEauthorrefmark{2}, Anindya Iqbal\IEEEauthorrefmark{2} and
Gias Uddin\IEEEauthorrefmark{3}}
\IEEEauthorblockA{\IEEEauthorrefmark{1}Ahsanullah University of Science and Technology, Dhaka, Bangladesh\\
\IEEEauthorrefmark{2}Bangladesh University of Engineering and Technology, Dhaka, Bangladesh\\
\IEEEauthorrefmark{3}York University, Canada\\
Email: \IEEEauthorrefmark{1}sshibli745@gmail.com,
\IEEEauthorrefmark{2}tahmidhasan@cse.buet.ac.bd,
\IEEEauthorrefmark{2}anindya@cse.buet.ac.bd,
\IEEEauthorrefmark{3}guddin@yorku.ca}}


\maketitle
\balance
\begin{abstract}
We present a novel approach - CLAA - for API aspect detection in API reviews that utilizes transformer models trained with a supervised contrastive loss objective function. We evaluate CLAA using performance and impact analysis. For performance analysis, we utilized a benchmark dataset on developer discussions collected from Stack Overflow and compare the results to those obtained using state-of-the-art transformer models. Our experiments show that contrastive learning can significantly improve the performance of transformer models in detecting aspects such as Performance, Security, Usability, and Documentation. 
For impact analysis, we performed empirical and developer study. On a randomly selected and manually labeled 200 online reviews, CLAA achieved 92\% accuracy while the SOTA baseline achieved 81.5\%. According to our developer study involving 10 participants, the use of \textit{Stack Overflow + CLAA} resulted in increased accuracy and confidence during API selection. Replication package: \url{https://github.com/disa-lab/Contrastive-Learning-API-Aspect-ASE2023}.
\end{abstract}


\section{Introduction}
API (Application Programming Interface) review is a crucial  process in software engineering that builds insight on an API's functionality, performance, and overall quality. Feedback from API users is essential in this process as it helps to identify areas for improvement and leads to better software development process \cite{uddin_understand}. By asking questions or sharing experiences and opinions about a particular API in online forums like Stack Overflow (SO), developers create a discussion thread that serves as a review \cite{uddin2019automatic}. Studies indicate that while there are multiple reviews available for a given API, developers tend to put higher importance on reviews that address specific aspects of an API \cite{yang2022aspect}. For instance, developers may be particularly interested in learning about an API's security features or its ease of use. These findings have led researchers to develop automated methods for accurately identifying the different aspects covered in API reviews \cite{uddin2017mining,uddin2019automatic,yang2022aspect,lin2019pattern, uddin2017opiner}.


Classifying API reviews based on predefined aspects is a challenging task. API reviews often use technical terms, domain-specific language, and jargon that are specific to programming languages, frameworks, and the API functionality. These terms may not be commonly used in other types of text and may not be present in the pre-trained language model's vocabulary, which makes it difficult for pre-trained language models to comprehend the meaning of the text. For example, the review \textit{"But this is also for transforming into well XMLs."} is related to "usability" aspect while the review \textit{"I'm searching the java library for parsing XML, I googled a bit but couldn't found other than dom4j"} is related to "community" aspect. 

\begin{figure}[h]
\centering
	\includegraphics[scale=0.38]{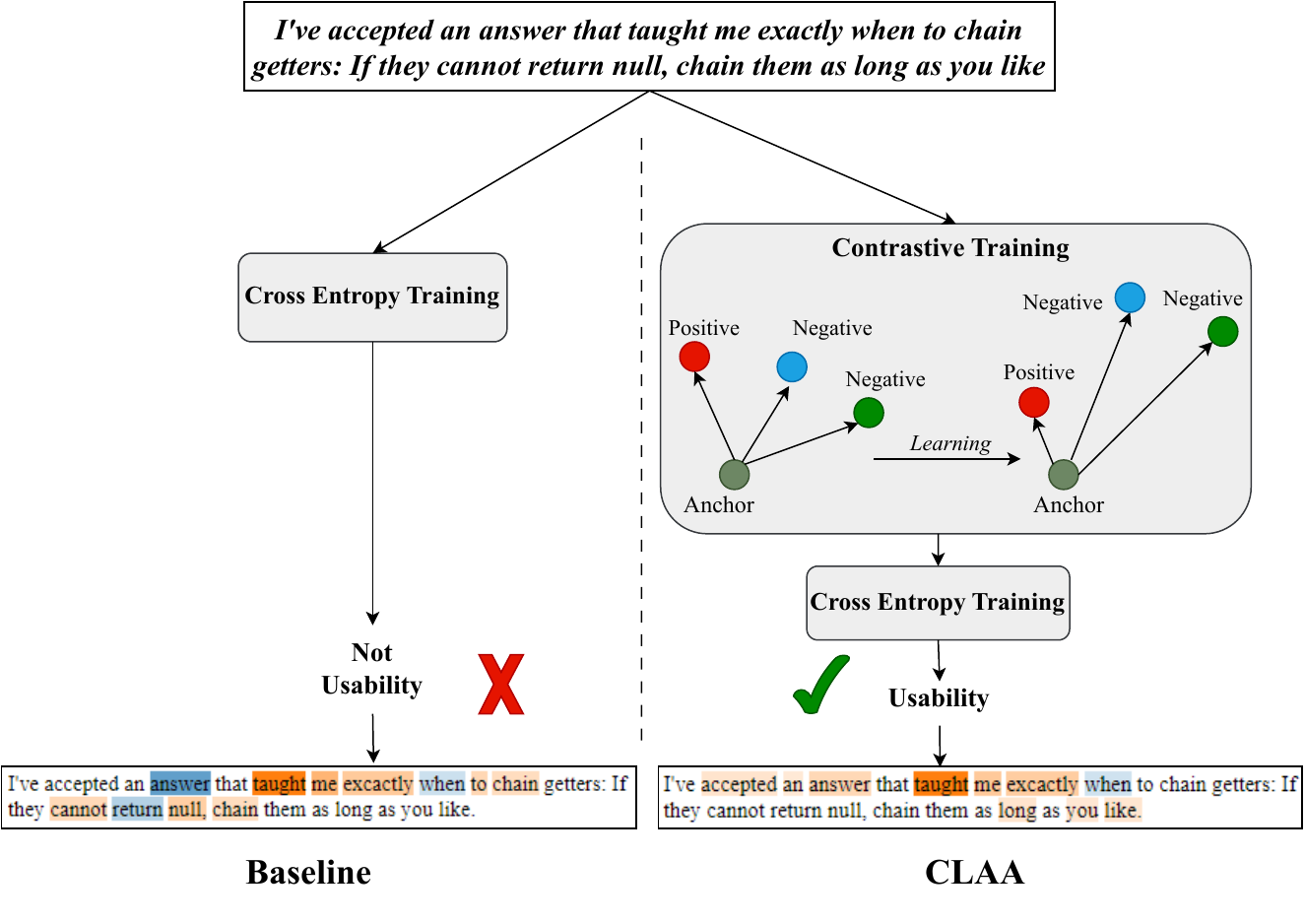}
\caption{An example showing how CLAA is superior to non-CL methods in our experiments. CL minimized the distance between an instance (anchor) and positive examples (instances from a certain aspect category) and maximized the distance between the anchor and negative examples (instances do not belong to that aspect). The output explanations are shown for both the baseline and CLAA with highlighted features. Orange color signifies features crucial for considering a text related to an aspect, while blue color represents the opposite. The lighter the color, the less importance the feature has.}
\label{fig:Motivation}
\end{figure}

In this paper, we propose a technique CLAA (Contrastive Learning for API Aspects). CLAA uses the principles of Contrastive Learning (CL) \cite{khosla2020supervised} and learns representations of API reviews that are specific to certain aspects, enabling the model to better distinguish between them. If a pre-trained language model is fine-tuned using CL, the model understands the technical vocabulary and domain-specific language used in API reviews better. Figure \ref{fig:Motivation} demonstrates an example of how CLAA can improve the API aspect detection task. We present a more clear visualization in Figure \ref{fig:visualization_graph} where we show the advantages of CLAA by highlighting the performance improvement over baselines. This approach utilizes supervised contrastive training objective to fine-tune each pre-trained transformer model (BERT \cite{devlin2018bert}, RoBERTa \cite{liu2019roberta}, XLNet \cite{yang2019xlnet}, ALBERT \cite{lan2019albert}, BERTOverflow \cite{tabassum2020code}, ELECTRA \cite{clark2020electra}, and T5 \cite{raffel2020exploring}) to learn the aspect-wise semantic representations of API review instances. These learned representations are then used to fine-tune the pre-trained transformers to act as sequence classifiers for API aspect detection task. To explain why the classifier makes a certain prediction, CLAA uses a perturbation based text explainer framework LIME (Local Interpretable Model-Agnostic Explanations) \cite{lime}. Unlike that of Yang et al. \cite{yang2022aspect}, our approach does not rely on only fine-tuning pre-trained transformers to act as a classifier. Rather CLAA uses two stage training: first the pre-trained transformers are fine-tuned to learn the representations of the reviews of different aspect categories through contrastive training, and then the transformers with knowledge of the aspect-wise learned representations are fine-tuned using CrossEntropy training.

CLAA holds the potential to offer a more insightful and contextually relevant API selection process which can enhance the decision making process in choosing specific APIs for specific projects. For example, consider a discussion regarding the \textit{usability} aspect of an API such as ``The new update introduced a more efficient API for data processing''. CLAA can differentiate that this type of text stands apart from non-aspect text like ``The team went out for lunch today''. By embedding these texts in a shared feature space, CLAA gains the ability to cluster aspect based API-related discussions in proximity while distancing them from conversations unrelated to a certain aspect. This ability to distinguish relevant from non-relevant texts contributes to the identification of crucial factors for API selection. It aligns recommendations with the project's prerequisites and customizes options based on user preferences, all of which can substantially enhance the API selection process for users.

We evaluate our proposed approach in two ways: (a) performance analysis and (b) impact analysis. For performance analysis, we compare CLAA with the state-of-the-art baselines through extensive experimentation. Experimental results show that it outperforms the SOTA results significantly. For impact analysis of CLAA, we conducted two experiments: (a) empirical study and (b) developer study. First, we collect all ``json'' and ``java'' related posts and comments from Stack Overflow. Then we apply the CLAA and the best performing baseline model (RoBERTa). 
 CLAA achieved 92\% accuracy by correctly predicting 184 sentences, while the baseline achieved 81.5\% accuracy by correctly predicting 163 sentences on randomly selected 200 reviews. Second, to demonstrate the usefulness of CLAA, we conduct a developer study in two phases. In the first phase, we ask 10 participants (five professional developers + five graduate level students) to perform two API selection related tasks using three different settings (\textit{Stack Overflow} only, \textit{Stack Overflow + baseline}, \textit{Stack Overflow + CLAA}). In the second phase, we collect their comparative feedback on the correctness, confidence and usefulness of CLAA. The findings of the study show that developers are more accurate about API selection when they are using CLAA.
In summary, we made the following contributions in this paper:
\begin{enumerate}
\item \textbf{CLAA}: We present CLAA, a novel approach for aspect analysis in API reviews. In CLAA, the aspect detection component leverages contrastive learning, which yields state-of-the-art results and the classifier output explanation component offers explanations to aid in aspect detection.
\item \textbf{Evaluation}: We conducted a thorough assessment of CLAA with both performance and impact analysis. It was compared with a closely related state-of-the-art approach \cite{yang2022aspect} and also a developer study involving ten participants was carried out.
\end{enumerate}

\section{CLAA: CL for API Aspects}
The CLAA tool comprises of two main components, i.e., the API aspect detection component and the outcome (of classifier) explanation component. The first component is responsible for classifying the API reviews into different aspect categories. The second component provides an interpretation of the predictions made by the first component. 

\subsection{API Aspect Detection Component}
We use API aspect detection component to categorize API reviews into some predefined aspects. The input to the component is an API related review, and the output is an aspect category. This component is composed of three parts: (1) contrastive training, (2) cross-entropy training, and (3) hyper-parameter tuning.


\subsubsection{Contrastive training} 
\begin{figure}[h]
\centering
	\includegraphics[width=1\linewidth]{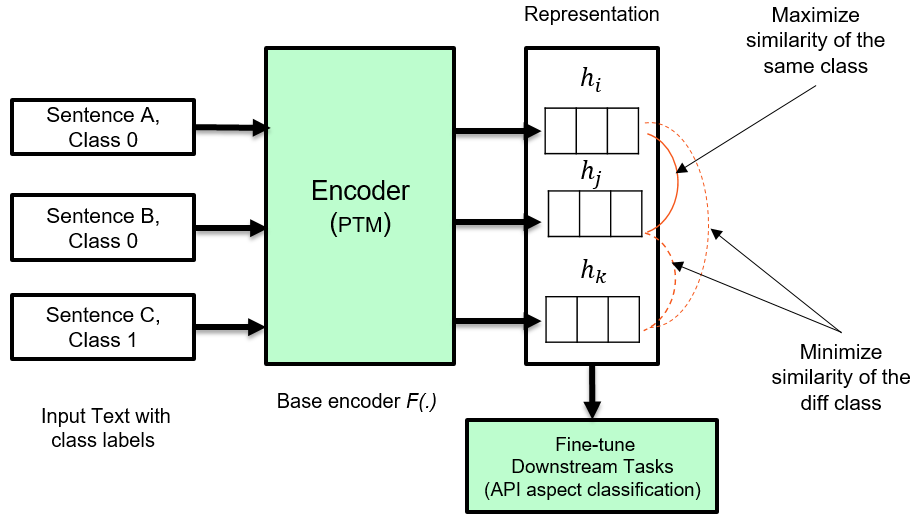}
\caption{Supervised contrastive training procedure.}
\label{fig:contrastive-image}
\end{figure}

Supervised contrastive learning is a technique widely used in Natural Language Processing (NLP) that requires training a model to differentiate between similar but distinct examples in order to improve its generalization ability \cite{gao2021simcse}. The basic idea behind the method is to train a model to identify the similarities and differences between two examples, with one example being the positive example and the other one being negative. Figure \ref{fig:contrastive-image} depicts a brief overview of the contrastive training procedure. At first, we separate the training data into two classes: instances belonging to a specific aspect are designated as the positive class, while the rest are considered negative. We split the training data into several batches of size $32$. For aspect categories with fewer than $100$ samples during training, to guarantee that each batch includes at least one positive class sample, the number of instances in these categories are doubled by duplicating each sentence. This data augmentation technique is considered minimal, as it is performed in a supervised setting and follows the unsupervised data augmentation method described in \cite{gao2021simcse}. The next step involved defining the model to be trained on the data. We utilize transformer models, as outlined in section \ref{PTMs}, as the encoder. The encoder learns the representations of the sentences in a batch. To train the encoder we use the NT-XENT (Normalized Temperature-Scaled Cross Entropy Loss) as the objective function which measures the similarity between the positive and negative examples in a pair and optimizes the model so that the similarity score between positive examples is maximized while the similarity between negative examples is minimized. The temperature scaling helps to balance the trade-off between the two objectives and adjust the sharpness of the optimization process. 
We use the supervised contrastive loss mentioned in Khosla et al. \cite{khosla2020supervised}, which can handle multiple positive and negative instances in a single batch. The goal of this loss function is to increase the similarity between positive instances and decrease the similarity between negative instances within a batch. We utilize the learned representations of the encoder to fine-tune the transformers for classifying API reviews. This process is repeated for each aspect and each transformer.

\subsubsection{Cross-entropy training} We fine-tune the transformer encoder as a sequence classifier to categorize API reviews. We employ \textit{binary cross-entropy loss}, a commonly used loss function in binary classification during fine-tuning. Given a predicted probability distribution $y_{\text{pred}}$ and the true label $y_{\text{true}}$, the binary cross-entropy loss is calculated as follows:
\begin{equation}
    \text{BCEloss} = -(y_{\text{true}} * \log(y_{\text{pred}}) + (1-y_{\text{true}}) * \log(1-y_{\text{pred}}))
\end{equation}
The objective is to minimize this loss, i.e. to make the predicted probability $y_{\text{pred}}$ as close as possible to the true label $y_{\text{true}}$.

\subsubsection{Hyper-parameter settings} For contrastive training, we used $0.063$ and $0.1$ as the temperature in NT-XENT loss function. We fine-tuned transformers for $5$ epochs with a batch size of $32$, \textit{AdamW} as optimizer, and a learning rate of $5e-05$. To have at least one positive sample per batch, we doubled the sample size for aspects with fewer than $100$ samples. 
For cross-entropy training, we fine-tuned transformers to act as classifiers by adding a dropout and linear layer and a varying learning rates of $1e-05, 2e-05$, and $5e-05$. We used $10-fold$ cross-validation, with a maximum sequence length of $160$ and \textit{binary cross-entropy} loss. We ran the experiments on Google Colab \cite{bisong2019google} and the pre-trained models were implemented using Hugging Face Transformer library.

\subsection{Classifier Output Explanation Component}
We utilize LIME (Local Interpretable Model-Agnostic Explanations) \cite{lime}, a widely used text explanation framework that assists in comprehending how the classifier in API aspect detection component of CLAA performs predictions. For instance, the sentence ``\textit{CBC encryption in itself is not thread self}" is a review regarding security aspects, where the words ``\textit{encryption}" and ``\textit{thread}" are crucial to determine it as a security aspect. LIME can determine the words or phrases in a review that significantly influenced the decision of the classifier in CLAA. LIME selects a specific piece of text that the machine learning model has made a prediction on. It then creates a number of perturbed versions of the text by making small, random changes to the original text. These perturbed versions are used to approximate the local behavior of the machine learning model around the original text. LIME uses a simpler, interpretable model (such as a logistic regression model) to make predictions on the perturbed versions of the text. The goal is to create a model that can accurately predict how the original machine learning model would behave around the original text, but in a way that is easier to understand. LIME then examines the simpler model to determine which words or phrases were most important in predicting the classifier's output for the original text. This can help explain why the classifier made the prediction. 


\section{Evaluation}
We answer two research questions:
\begin{enumerate}
    \item \textbf{Performance Analysis.} How accurate is CLAA to detect API aspects?
    \item \textbf{Impact Analysis.} How effective is CLAA to support users during their analysis of API reviews?
 
\end{enumerate}

\subsection{Study Setup for Performance Analysis}\label{dataset}
We used the dataset from Uddin et al. \cite{uddin2017opiner} for performance comparison. The dataset consists of $4,522$ sentences extracted from $1,338$ Stack Overflow (SO) posts and was manually labeled. Out of the $4,522$ sentences, $4,307$ belong to a single API aspect, $209$ belong to two aspects, and the remaining $6$ belong to more than two aspects. The data distribution of the benchmark dataset is presented in Table \ref{tab:distribution}. 

\begin{table}[h]
\centering
\caption{The distribution of the API review aspects along with definition in Opiner \cite{uddin2017opiner} dataset.}
\label{tab:distribution}
\resizebox{\columnwidth}{!}{%
\begin{tabular}{|c|c|c|}
\hline
\textbf{Aspects} & \textbf{Definition}                                                                                                                                              & \textbf{\begin{tabular}[c]{@{}c@{}}No. of \\ Samples\end{tabular}} \\ \hline
Performance      & \begin{tabular}[c]{@{}c@{}}Facilitates the comparison between two or more \\ APIs in terms of performance, resource consumption.\end{tabular}                    & 348 (7.7\%)                                                        \\ \hline
Usability        & \begin{tabular}[c]{@{}c@{}}Discussion regarding the API usage, applications \\ and integration challenges.\end{tabular}                                          & 1437 (31.8\%)                                                      \\ \hline
Security         & \begin{tabular}[c]{@{}c@{}}Discussion related to the level of data \\ security provided by the API.\end{tabular}                                                 & 163 (3.6\%)                                                        \\ \hline
Community        & \begin{tabular}[c]{@{}c@{}}Discussion related to the activity level of \\ the community of API practitioners.\end{tabular}                                       & 93 (2.1\%)                                                         \\ \hline
Compatibility    & \begin{tabular}[c]{@{}c@{}}Discussion related to the API's compatibility \\ with the specified framework environments.\end{tabular}                              & 93 (2.1\%)                                                         \\ \hline
Portability      & \begin{tabular}[c]{@{}c@{}}Discussion related to the adaptability of the API in \\ circumstances such as numerous \\ operating system environments.\end{tabular} & 70 (1.5\%)                                                         \\ \hline
Documentation    & \begin{tabular}[c]{@{}c@{}}Discussion related to the clarity and completeness\\  of the API's official documentation.\end{tabular}                               & 256 (5.6\%)                                                        \\ \hline
Bug              & \begin{tabular}[c]{@{}c@{}}Discussion related to the overall existence \\ or absence of bugs and faults in the API.\end{tabular}                                 & 189 (4.2\%)                                                        \\ \hline
Legal            & \begin{tabular}[c]{@{}c@{}}Discussion related to the level of legal authorization\\  and access provided for API usage.\end{tabular}                             & 50 (1.1\%)                                                         \\ \hline
OnlySentiment    & \begin{tabular}[c]{@{}c@{}}Discussion that expresses simply opinion \\ regarding an API, with no technical details.\end{tabular}                                 & 348 (7.7\%)                                                        \\ \hline
Others           & \begin{tabular}[c]{@{}c@{}}Discussion related to the APIs that do not \\ fall under the previously described aspects.\end{tabular}                               & 1699 (37.6\%)                                                      \\ \hline
\end{tabular}%
}
\end{table}

\subsubsection{Models used to detect aspects in CLAA}\label{PTMs}
We utilized seven pre-trained transformer models in the API aspect detection component of CLAA. Firstly, we employed BERT \cite{devlin2018bert}, which was trained using `Masked Language Modeling' and `Next Sentence Prediction' objectives. Another model we used was RoBERTa \cite{liu2019roberta}, a modified version of BERT that incorporated a larger training dataset and different training strategies. Additionally, we employed ALBERT \cite{lan2019albert}, which maintained BERT's performance while reducing its parameters. For domain-specific tasks, we utilized BERTOverflow \cite{tabassum2020code}, trained with a large dataset from Stack Overflow. We also incorporated XLNet \cite{yang2019xlnet}, a generalized auto-regressive model that considers inter-dependency between tokens during permutation language modeling. Another model we used was ELECTRA \cite{clark2020electra}, which focused on identifying replaced tokens in the input sequence. Lastly, we included T5 \cite{raffel2020exploring} , trained with a `Masked Language Modeling' goal, but with a different approach to handling consecutive tokens. Table \ref{tab:version} describes the architecture details.

\begin{table}[htbp]
\centering
\caption{Architecture details of of pre-trained transformer models (L = Layers, H = Heads, P = Parameters).}
\label{tab:version}
\resizebox{\columnwidth}{!}
{\begin{tabular}{@{}|c|c|c|c|c|@{}}
\toprule
\textbf{Architecture} & \textbf{Used Model}               & \textbf{L} & \textbf{H} & \textbf{P} \\ \midrule
BERT                  & bert-base-uncased                 & 12         & 12         & 110M       \\ \midrule
RoBERTa               & roberta-base                      & 12         & 12         & 123M       \\ \midrule
BERTOverflow          & jeniya/BERTOverflow               & 12         & 12         & 149M       \\ \midrule
ALBERT                & albert-base-v2                    & 12         & 12         & 11M        \\ \midrule
XLNet                 & xlnet-base-cased                  & 12         & 12         & 110M       \\ \midrule
ELECTRA               & google/electra-base-discriminator & 12         & 12         & 110M       \\ \midrule
T5                    & t5-small                          & 6          & 8          & 60M        \\ \bottomrule
\end{tabular}}
\end{table}

\begin{table*}
\centering
\caption{Performance comparison between CLAA and the baselines (fine-tuned transformers without contrastive learning). Here, all the results are averaged across all the aspects. STDV is the standard deviation over F1-score.}
\label{tab:perf-comp}
\begin{tabular}{|c|c|c|c|c|c|c|c|c|c|c|c|} 
\cline{2-12}
\multicolumn{1}{c|}{} & \multicolumn{4}{c|}{\textbf{CLAA}}                                                  & \multicolumn{4}{c|}{\textbf{Baseline}}                                              & \multicolumn{3}{c|}{\textbf{Improvement by CLAA (\%)}}  \\ 
\hline
\textbf{Model}        & \textbf{~F1~} & \textbf{STDV} & \textbf{~MCC~} & \textbf{~AUC~} & \textbf{~F1~} & \textbf{STDV} & \textbf{~MCC~} & \textbf{~AUC~} & \textbf{~F1~} & \textbf{~MCC~} & \textbf{~AUC~}             \\ 
\hline
BERT                  & \textbf{0.9745}     & 0.010             & 0.7684             & 0.8815               & 0.9422              & 0.008             & 0.6068             & 0.7850               & 3.2316            & 16.1630          & 9.6450                         \\ 
\hline
RoBERTa               & \textbf{0.9776}     & 0.012             & 0.7620             & 0.8831               & 0.9429              & 0.007             & 0.5846             & 0.7748               & 3.4700            & 17.7400          & 10.8300                        \\ 
\hline
Bert Overflow         & \textbf{0.9107}     & 0.011             & 0.1415             & 0.5650               & 0.9076              & 0.011             & 0.1337             & 0.5587               & 0.3080            & 0.7770           & 0.6300                         \\ 
\hline
XLNET                 & \textbf{0.9448}     & 0.012             & 0.5194             & 0.7488               & 0.9380              & 0.011             & 0.5197             & 0.7451               & 0.6820            & -0.0300          & 0.3700                         \\ 
\hline
ALBERT                & \textbf{0.9725}     & 0.009             & 0.6757             & 0.8259               & 0.9306              & 0.012             & 0.4720             & 0.7142               & 4.1910            & 20.3760          & 11.1740                        \\ 
\hline
ELECTRA               & \textbf{0.9677}     & 0.007             & 0.5368             & 0.7726               & 0.9350              & 0.018             & 0.4920             & 0.7361               & 3.2650            & 4.4770           & 3.6500                         \\ 
\hline
T5                    & \textbf{0.9055}     & 0.009             & 0.2187             & 0.5707               & 0.9033              & 0.009             & 0.1886             & 0.5594               & 0.2200            & 3.0100           & 1.1250                         \\
\hline
\end{tabular}
\end{table*}



\subsubsection{Evaluation metrics}
We evaluated the performance of each classifier inside CLAA using five metrics: \textit{weighted precision (P), weighted recall (R), weighted f1 score (F1), matthews correlation coefficient (MCC),} and \textit{weighted area under ROC curve (ROC AUC)}. We considered F1 score as the primary evaluation metric. To validate the results of the performance comparison, we conducted a \textit{paired bootstrap re-sampling test} \cite{koehn2004statistical}. This test allows us to assess the significance of the differences between the classifiers in binary classification by repeatedly re-sampling the original data and comparing the accuracy and F1-score of the classifiers. 

\subsection{Study Setup for Impact Analysis}
We conducted two experiments: 

\subsubsection{Empirical} We utilized the March 2023 Stack Overflow data dump, which was the latest available at the time. We collected posts and comments using two tags: ``json''and ``java''. Our experimental dataset comprised all the posts and comments labeled with at least one of these tags, amounting to a total of $8,722$ posts and $12,483$ comments. Since our focus was on textual analysis, we used the \textit{BeautifulSoup} library to extract only relevant information such as titles and URLs. Additionally, we employed the NLTK sentence tokenizer to extract all sentences from the posts and comments, which resulted in approximately $86,853$ sentences. To evaluate the performance of both the best performing baseline model and CLAA, this dataset was used. Each tool generated a list of sentences labeled as an aspect. We then randomly selected a subset of sentences from the dataset and manually compared the accuracy of each model on those sentences.

\subsubsection{User Study} To analyze the effectiveness of CLAA in assisting development tasks, we conducted a user study by following the work of Uddin et al. \cite{uddin2017automatic}.
Ten developers took part in the study, and each of them completed two tasks involving the selection of an API from a pool of two options. Once the tasks were completed, the developers were invited to provide feedback on their experience of using CLAA through an online data collection tool.

\noindent \textbf{A. Tasks.} We created two tasks for the study following that of Uddin et al. \cite{uddin2017automatic}. Each task used a different set of APIs. Both tasks involved choosing an API from a pool of two options. The first set of options included "GSON" and "org.json", while the second set included "jackson" and "json-lib".

\textbf{(T1)} The participants were tasked to choose between two APIs, GSON and org.json, based on two criteria: a) usability, and b) licensing usage. The correct answer was GSON.

\textbf{(T2)} The study required participants to choose between two APIs, Jackson and json-lib, based on two criteria: a) performance, and b) pre-installed base in leading frameworks. The correct answer was Jackson.
\noindent For each task, we asked each developer to complete it in three different settings:\\
(1) \textbf{SO only}: Using Stack Overflow as the only resource.\\
(2) \textbf{SO + baseline}: Using both Stack Overflow and the best performing baseline (RoBERTa).\\
(3) \textbf{SO + CLAA}: Using both Stack Overflow and CLAA.
\noindent For each task and setting, each developer was required to provide the following answers: (1) \textbf{Selection}: The API they chose. (2) \textbf{Confidence}: Their level of confidence while making the selection, measured on a five-point scale ranging from fully confident (value 5) to fully unsure (value 1). (3) \textbf{Reasoning}: The reason or reasons for their selection, expressed in one or more sentences. Once the developer finished the assigned tasks, we requested their feedback regarding their experience with CLAA. Specifically, we inquired about the extent to which they would utilize CLAA for future development tasks and also invited them to suggest improvements that could be made to CLAA. The developers were encouraged to provide detailed responses to these questions.

\noindent \textbf{B. Participants.}
We recruited a total of ten developers for our study, with experience levels ranging from 1 to 12 years. Of the participants, five were graduate students and the remaining five were professional developers. Three of the professional developers were affiliated with a software company, while the remaining two were recruited through Freelancer.com. To ensure that each participant fully understood the study and its requirements, we contacted them directly via email, chat, and Skype. Each developer was granted access to our data collection tool. To facilitate the study, we developed inference scripts for both CLAA and the baseline with the highest performance. These scripts are designed to take API reviews as input and generate an aspect as output. We made these available online, allowing each study participant to utilize them for each task.

\section{Performance Analysis of CLAA (RQ1)}
We investigate the following sub-RQs:\\
1. Does CLAA offer improvement over the baselines?\\
2. What are the misclassification categories observed in the predictions of both the baseline and CLAA?

\subsection{RQ1.1 Does CLAA offer improvement over the baselines?}
We answer the following sub-RQs to answer this research question:
\begin{enumerate}[label=(\alph*)]
\item Does contrastive learning improve the performance of pre-trained models compared to the state-of-the-art performance of transformer based models?
\item What is the impact of employing different contrastive learning objectives on the performance of CLAA?
\end{enumerate}
\vspace{0.5em}
\noindent \textit{1) RQ1.1a Does contrastive learning improve the performance of pre-trained
models compared to the state-of-the-art performance of transformer based models?}

\textit{\underline{Approach.}} To perform aspect-wise binary classification on API reviews, we divide the dataset into two categories: reviews that are related to a specific aspect, and reviews that are not. We fine-tune the pre-trained transformers using only cross-entropy training to establish a baseline performance. After that, we apply CLAA to the same dataset to determine if it can enhance the performance.

\begin{figure*}[t]
		\centering
		\begin{subfigure}[b]{.5\columnwidth}
		    \centering
			\includegraphics[width=1\linewidth]{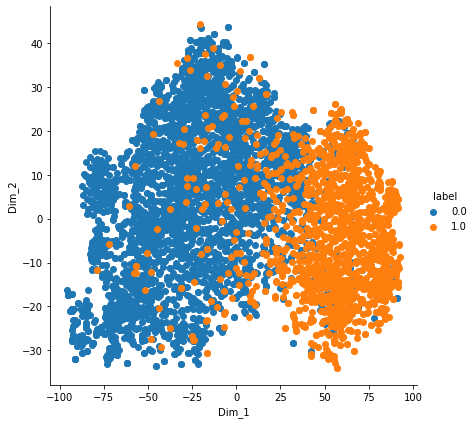}
                 \subcaption{Usability without CLAA}
                \label{fig:Usa_before}
		\end{subfigure}
		\begin{subfigure}[b]{.5\columnwidth}
		    \centering
			\includegraphics[width=1\linewidth]{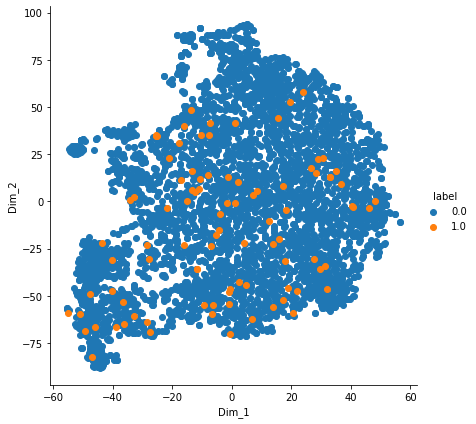}
                \subcaption{Compatibility without CLAA}
                \label{fig:Comp_before}
		\end{subfigure}
        \begin{subfigure}[b]{.5\columnwidth}
		    \centering
			\includegraphics[width=1\linewidth]{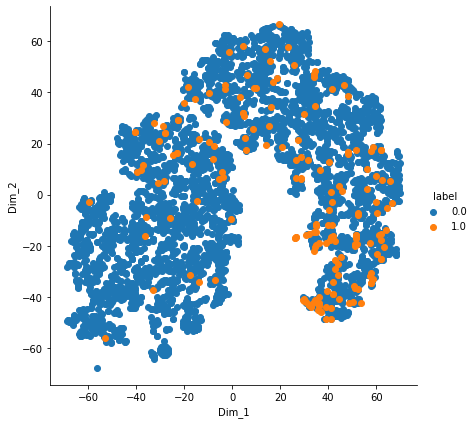}
			\subcaption{Bug without CLAA}
                \label{fig:Bug_before}
		\end{subfigure}
		\begin{subfigure}[b]{.5\columnwidth}
		    \centering
			\includegraphics[width=1\linewidth]{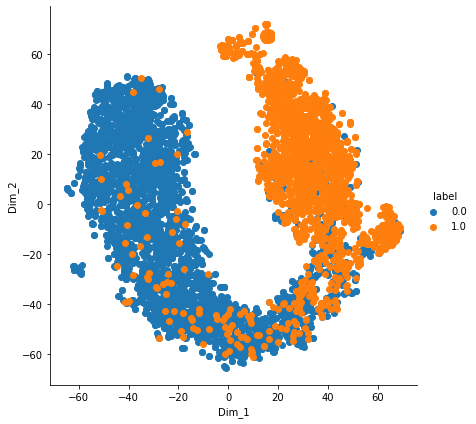}
                \subcaption{Others without CLAA}
                \label{fig:Others_before}
		\end{subfigure}\\   
        \begin{subfigure}[b]{.5\columnwidth}
		    \centering
			\includegraphics[width=1\linewidth]{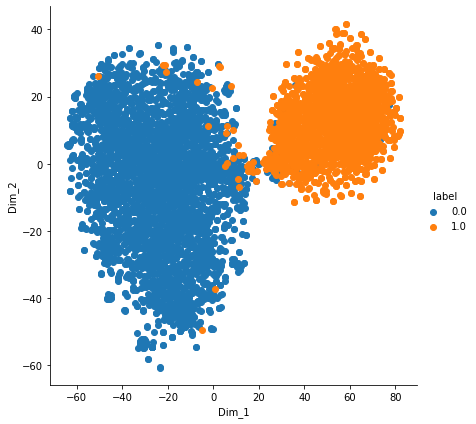}
                 \subcaption{Usability with CLAA}
                \label{fig:Usa_after}
		\end{subfigure}
		\begin{subfigure}[b]{.5\columnwidth}
		    \centering
			\includegraphics[width=1\linewidth]{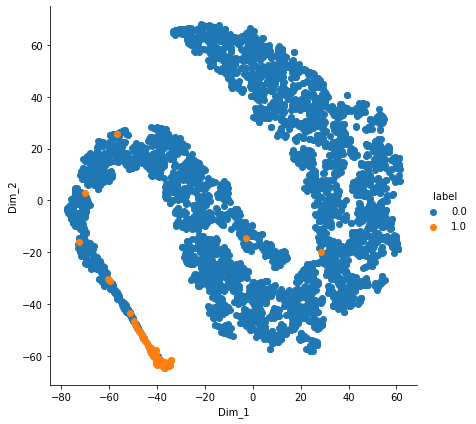}
                \subcaption{Compatibility with CLAA}
                \label{fig:Comp_after}
		\end{subfigure}
        \begin{subfigure}[b]{.5\columnwidth}
		    \centering
			\includegraphics[width=1\linewidth]{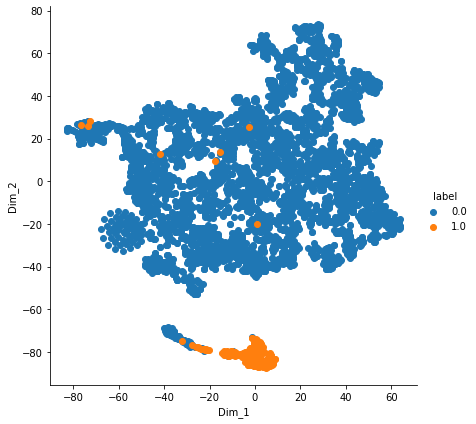}
			\subcaption{Bug with CLAA}
                \label{fig:Bug_after}
		\end{subfigure}
		\begin{subfigure}[b]{.5\columnwidth}
		    \centering
			\includegraphics[width=1\linewidth]{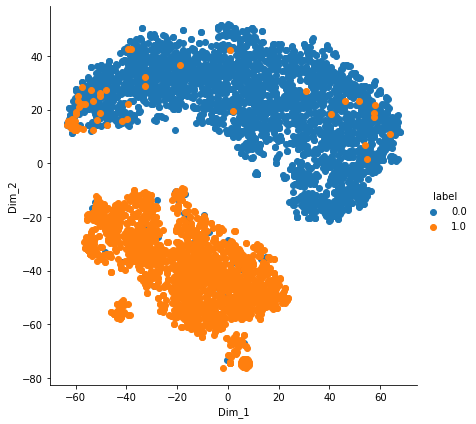}
                \subcaption{Others with CLAA}
                \label{fig:Others_after}
		\end{subfigure}
		\caption{Visualization of the embeddings produced by ALBERT pre-trained model on four API aspects: \textit{usability, compatibility, bug, others}. Sub-figures \textit{(a to d)} presents the embeddings produced as a baseline. Sub-figures \textit{(e to h)} represents the embeddings produced by CLAA which exhibit a clear decision boundary for binary API aspect detection task.}
		\label{fig:visualization_graph}
\end{figure*}

\textit{\underline{Results.}} 
To evaluate CLAA, we have used seven pre-trained transformer models. The average performance of these models on different aspects, including performance improvement, is presented in Table \ref{tab:perf-comp}. The fine-tuning with contrastive learning in CLAA led to a significant improvement to the average performance of the transformer models, except for BERTOverflow, XLNet, and T5. Out of all the transformer models used in CLAA, RoBERTa demonstrates the highest performance in terms of average F1, MCC, and AUC scores, despite having a similar average F1 score to BERT and ALBERT. However, fine-tuning BERTOverflow, XLNet, and T5 with contrastive learning does not have a significant impact on their performance. Furthermore, XLNet without contrastive learning exhibits a better average MCC score.
Due to space limitation, we report the detailed performance comparison on each of the 11 aspects in the Github repository. According to the results, BERT demonstrated impressive performance for 9 of the 11 aspects (excluding Community and Compatibility) with high MCC, AUC, and F1 scores. On the other hand, RoBERTa displayed poor performance in the Compatibility aspect with MCC and AUC scores of 0 and 0.5, respectively, but performed better in other aspects, exhibiting MCC scores higher than 47\% which indicates a very strong positive correlation and AUC scores higher than 89\% which mark it as a decent classifier. Additionally, in the Community aspect, RoBERTa with contrastive learning achieved the best performance, with an MCC score of 87.87\% and an AUC score of 91.86\%. BERTOverflow had poor performance on six aspects, including Security, Community, Compatibility, Portability, Bug, and Legal, as evidenced by a MCC score of 0 and an AUC score of 0.5, indicating no correlation and random guessing. Overall, BERTOverflow performed poorly and had little predictive ability on most aspects. Although XLNet and T5 showed good evaluation scores in API aspect detection, the impact of contrastive training on their performance was minimal. Even though their average F1 score was improved, their MCC and AUC scores were unsatisfactory, indicating that they were not effective in correctly identifying the API aspects. ALBERT demonstrated the most significant average F1 score improvement among all models. However, it had poor performance in the Compatibility aspect with low MCC and AUC scores. Nonetheless, for all other aspects, ALBERT achieved an MCC score of over 73\% and an AUC score of over 83\%. Similarly, ELECTRA also showed significant improvement in performance, but both models performed poorly in the Legal aspect, with an MCC score of 0 and an AUC score of 0.5, indicating random guessing. To clarify the improvement brought about by CLAA over the baseline models, we utilized t-Distributed Stochastic Neighbor Embedding (t-SNE) \cite{van2008visualizing} to create visual representations of the sentence embeddings produced by the pre-trained transformer models. 
In Figure \ref{fig:visualization_graph}, we present a visualization of sentence embeddings generated by the ALBERT transformer model in both the CLAA and baseline method. We specifically chose the ALBERT model since it demonstrated the greatest improvement in terms of F1-score within the CLAA framework. In the upper images (\ref{fig:Usa_before} to \ref{fig:Others_before}), we can observe the embeddings of two classes (aspect and non-aspect) generated by the baseline ALBERT model. The orange and blue dots represent embedding vectors in a two-dimensional space, where orange indicates aspect reviews and blue indicates non-aspect reviews. The visualization shows that there is a significant overlap between the embeddings of the two classes, and the decision boundary between them is not clear. This implies that it is challenging to separate the two classes accurately using the baseline ALBERT model. 
The lower images (\ref{fig:Usa_after} to \ref{fig:Others_after}) in Figure \ref{fig:visualization_graph} illustrate the embeddings of the two classes produced by CLAA. The visualization reveals dense clusters of embedding vectors that are clearly separated by a relatively distinct margin between the two clusters, indicating why CLAA has achieved significant performance improvement.

\vspace{0.5em}

\begin{table}
\centering
\caption{Performance comparison of BERT and RoBERTa in CLAA with three different contrastive learning objectives.}
\label{tab:compare-losses}
\resizebox{\columnwidth}{!}
{\begin{tabular}{ccccccc} 
\cline{2-7}
  & \multicolumn{3}{c}{\textbf{BERT}}       & \multicolumn{3}{c}{\textbf{RoBERTa}}                                                                                                                                                                            \\ 
\hline
\begin{tabular}[c]{@{}c@{}}\textbf{Objective}\\\textbf{Function}\end{tabular} & \begin{tabular}[c]{@{}c@{}}\textbf{Avg}\\\textbf{ F1}\end{tabular} & \begin{tabular}[c]{@{}c@{}}\textbf{Avg}\\\textbf{ MCC}\end{tabular} & \begin{tabular}[c]{@{}c@{}}\textbf{Avg~}\\\textbf{AUC}\end{tabular} & \begin{tabular}[c]{@{}c@{}}\textbf{Avg}\\\textbf{ F1}\end{tabular} & \begin{tabular}[c]{@{}c@{}}\textbf{Avg}\\\textbf{ MCC}\end{tabular} & \begin{tabular}[c]{@{}c@{}}\textbf{Avg}\\\textbf{ AUC}\end{tabular}  \\ 
\hline
NT-XENT loss                                                                  & 0.974                                                              & 0.768                                                               & 0.881                                                               & \textbf{0.978}                                                     & \textbf{0.762}                                                      & \textbf{0.883}                                                       \\
TripletMarginLoss                                                             & \textbf{0.988}                                                     & \textbf{0.880}                                                      & \textbf{0.932}                                                      & 0.950                                                              & 0.683                                                               & 0.847                                                                \\
ContrastiveLoss                                                               & 0.972                                                              & 0.813                                                               & 0.900                                                               & 0.944                                                              & 0.377                                                               & 0.687                                                                \\
\hline
\end{tabular}}
\end{table}
\noindent \textit{2) RQ1.1b What is the impact of employing different contrastive learning objectives on the performance of CLAA?}

\textit{\underline{Approach.}} We utilized two distinct supervised contrastive learning objectives, namely \textit{TripletMarginLoss} \cite{conneau2017supervised} and \textit{ContrastiveLoss} \cite{chopra2005learning}, to fine-tune BERT and RoBERTa in CLAA. The supervised version of \textit{TripletMarginLoss} and \textit{ContrastiveLoss} take into account labeled data for calculating loss, maximizing similarity between positive samples (same class) and minimizing similarity between negative samples (different classes). BERT fine-tuned with the \textit{TripletMarginLoss} training objective, displayed a significant improvement of approximately 3-5\% in average F1-Score across all aspects, compared to all the baseline models. This inspires us to explore whether this performance enhancement remains consistent with different training objectives.


\textit{\underline{Results.}} 
We compare BERT and RoBERTa's performance on all aspects using NT-XENT loss, TripletMarginLoss and ContrastiveLoss. BERT fine-tuned with TripletMarginLoss outperforms NT-XENT loss and ContrastiveLoss. For example, in the "Usability" aspect, BERT fine-tuned with NT-XENT loss achieved an F1-score of 0.86, while BERT fine-tuned with TripletMarginLoss achieved 0.98, and ContrastiveLoss achieved 0.91. In contrast, for the "Others" aspect, BERT fine-tuned with ContrastiveLoss has a poor F1-score of 0.91, while BERT fine-tuned with NT-XENT and TripletMarginLoss achieved around 0.98. Conversely, for RoBERTa, the results are the opposite. RoBERTa fine-tuned with TripletMarginLoss performs poorly with an F1-score of only 0.71, while RoBERTa fine-tuned with NT-XENT loss performs best with an F1-score of 0.94. RoBERTa fine-tuned with ContrastiveLoss achieved an F1-score of 0.85. Additionally, we observe a decrease in performance for RoBERTa fine-tuned with TripletMarginLoss and ContrastiveLoss for the Bug and Others aspects. Table \ref{tab:compare-losses} demonstrates that BERT with TripletMarginLoss outperforms NT-XENT loss in terms of average F1, MCC, and AUC scores by a margin of 1.4\%, 11\%, and 5\%, respectively. BERT, fine-tuned with TripletMarginLoss, exhibits the best performance among all transformer models for API aspect detection. 

\begin{table}
\centering
\caption{Statistics on how the misclassification of a tool can be corrected by another tool.}
\label{tab:error-comp}
\resizebox{\columnwidth}{!}
{\begin{tabular}{clccc} 
\hline
\multirow{2}{*}{\textbf{Aspect}}                                         & \multicolumn{1}{c}{\multirow{2}{*}{\begin{tabular}[c]{@{}c@{}}\textbf{Tool }\\\textbf{wrong}\end{tabular}}} & \multirow{2}{*}{\begin{tabular}[c]{@{}c@{}}\textbf{\# of misclassified }\\\textbf{intances}\end{tabular}} & \multicolumn{2}{c}{\begin{tabular}[c]{@{}c@{}}\textbf{Tools could correct }\\\textbf{wrong aspect}\end{tabular}}  \\ 
\cline{4-5}
                                                                         & \multicolumn{1}{c}{}                                                                                        &                                                                                                           & \textbf{Baseline} & \textbf{CLAA}                                                                                 \\ 
\hline
\multirow{3}{*}{Performance}                                             & \textbf{Baseline~~}                                                                                         & 339                                                                                                       &                   & 89\%                                                                                          \\
                                                                         & \textbf{CLAA}                                                                                               & 73                                                                                                        & 48\%              &                                                                                               \\
                                                                         & \textbf{Both}                                                                                               & 38                                                                                                        &                   &                                                                                               \\ 
\hline
\multirow{3}{*}{Usability}                                               & \textbf{Baseline}                                                                                           & 999                                                                                                       &                   & 94\%                                                                                          \\
                                                                         & \textbf{CLAA}                                                                                               & 307                                                                                                       & 81\%              &                                                                                               \\
                                                                         & \textbf{Both}                                                                                               & 59                                                                                                        &                   &                                                                                               \\ 
\hline
\multirow{3}{*}{Security}                                                & \textbf{Baseline}                                                                                           & 113                                                                                                       &                   & 81\%                                                                                          \\
                                                                         & \textbf{CLAA}                                                                                               & 95                                                                                                        & 77\%              &                                                                                               \\
                                                                         & \textbf{Both}                                                                                               & 22                                                                                                        &                   &                                                                                               \\ 
\hline
\multirow{3}{*}{Community}                                               & \textbf{Baseline}                                                                                           & 136                                                                                                       &                   & 65\%                                                                                          \\
                                                                         & \textbf{CLAA}                                                                                               & 50                                                                                                        & 4\%               &                                                                                               \\
                                                                         & \textbf{Both}                                                                                               & 48                                                                                                        &                   &                                                                                               \\ 
\hline
\multirow{3}{*}{Compatibility}                                           & \textbf{Baseline}                                                                                           & 140                                                                                                       &                   & 34\%                                                                                          \\
                                                                         & \textbf{CLAA}                                                                                               & 135                                                                                                       & 31\%              &                                                                                               \\
                                                                         & \textbf{Both}                                                                                               & 93                                                                                                        &                   &                                                                                               \\ 
\hline
\multirow{3}{*}{Portability}                                             & \textbf{Baseline}                                                                                           & 77                                                                                                        &                   & 83\%                                                                                          \\
                                                                         & \textbf{CLAA}                                                                                               & 14                                                                                                        & 7\%               &                                                                                               \\
                                                                         & \textbf{Both}                                                                                               & 13                                                                                                        &                   &                                                                                               \\ 
\hline
\multirow{3}{*}{Documentation~}                                          & \textbf{Baseline}                                                                                           & 235                                                                                                       &                   & 80\%                                                                                          \\
                                                                         & \textbf{CLAA}                                                                                               & 81                                                                                                        & 42\%              &                                                                                               \\
                                                                         & \textbf{Both}                                                                                               & 47                                                                                                        &                   &                                                                                               \\ 
\hline
\multirow{3}{*}{Bug}                                                     & \textbf{Baseline}                                                                                           & 163                                                                                                       &                   & 97\%                                                                                          \\
                                                                         & \textbf{CLAA}                                                                                               & 81                                                                                                        & 94\%              &                                                                                               \\
                                                                         & \textbf{Both}                                                                                               & 5                                                                                                         &                   &                                                                                               \\ 
\hline
\multirow{3}{*}{Legal}                                                   & \textbf{Baseline}                                                                                           & 63                                                                                                        &                   & 21\%                                                                                          \\
                                                                         & \textbf{CLAA}                                                                                               & 77                                                                                                        & 35\%              &                                                                                               \\
                                                                         & \textbf{Both}                                                                                               & 50                                                                                                        &                   &                                                                                               \\ 
\hline
\multirow{3}{*}{\begin{tabular}[c]{@{}c@{}}Only\\Sentiment\end{tabular}} & \textbf{Baseline}                                                                                           & 235                                                                                                       &                   & 65\%                                                                                          \\
                                                                         & \textbf{CLAA}                                                                                               & 118                                                                                                       & 31\%              &                                                                                               \\
                                                                         & \textbf{Both}                                                                                               & 82                                                                                                        &                   &                                                                                               \\ 
\hline
\multirow{3}{*}{Others}                                                  & \textbf{Baseline}                                                                                           & 954                                                                                                       &                   & 84\%                                                                                          \\
                                                                         & \textbf{CLAA}                                                                                               & 339                                                                                                       & 55\%              &                                                                                               \\
                                                                         & \textbf{Both}                                                                                               & 153                                                                                                       &                   &                                                                                               \\
\hline
\end{tabular}}
\end{table}

\subsection{RQ1.2 What are the misclassification categories observed in the predictions of both the baseline and CLAA?}
We answer the research question by considering the following two sub-RQs:
\begin{enumerate}[label=(\alph*)]
    \item How frequently can one tool correct the misclassification of another tool?
    \item What are the reasons behind the misclassification?
\end{enumerate}

\vspace{0.5em}
\noindent \textit{1) RQ1.2a How frequently can one tool correct the misclassification of another tool?}

\textit{\underline{Approach.}}
We identify all of the instances that were misclassified by both the baseline model and CLAA on an aspect-by-aspect basis. In both the models, we use ALBERT as the pre-trained model as it achieved the highest average F1 improvement. Then, we determine how often one tool can fix the misclassification of the other tool. Specifically, if one tool misclassifies an instance, the other can potentially correct it by predicting the correct API aspect.
Our study shows that the baseline and CLAA can work together to complement each other's strengths. Additionally, we report the frequency with which both models fail to correctly predict each aspect.

\textit{\underline{Results.}} Table \ref{tab:error-comp} shows the percentage of textual units that are misclassified by each tool and can be potentially corrected by another tool. The first column of the table represents the API aspects that each tool is designed to detect. The third column displays the number of API reviews that are misclassified by a particular tool mentioned in the second column. In the fourth and fifth columns, we have reported how often the other tools can correct the misclassifications made by the tool mentioned in the second column. Our analysis in Table \ref{tab:error-comp} reveals that CLAA performs better than the baseline model in correcting misclassifications for most API aspects, such as \textit{Bug, Usability, Performance, Portability}, and \textit{Community}. For instance, CLAA can correct a high percentage (97\%, 94\%, and 89\%) of misclassifications made by the baseline model for \textit{Bug, Usability}, and \textit{Performance} aspects, respectively. On the other hand, the baseline model can only correct a small percentage (7\% and 4\%) of misclassifications made by CLAA. However, the baseline model performs better than CLAA in correcting misclassifications for \textit{Legal} aspect, where it can correct 35\% of the misclassified instances compared to only 21\% corrected by CLAA. It is worth noting that both the baseline and CLAA struggle to detect certain aspects, such as \textit{Community, Compatibility, Legal}, and \textit{Others}, where they fail to classify a considerable number of API review instances i.e. 48, 93, 50, and 153, respectively. Nevertheless, our study shows that CLAA has a lower number of misclassified instances in general, indicating its effectiveness in the API aspect analysis task.

\vspace{0.3em}
\noindent \textit{2) RQ1.2b What are the reasons behind the misclassification?}

\textit{\underline{Approach.}}
To analyze the reasons behind misclassifications, we randomly selected 20 misclassified reviews from each of the eleven aspects, resulting in a total of 220 misclassified instances. We identified five error categories to analyze the reasons behind misclassifications, and we manually labeled the misclassified instances into one of these categories. The categories are: (a) general error, (b) politeness, (c) inability to deal with context information, (d) lack of domain-specific knowledge and (e) unknown tokens.

\begin{figure}[h]
\centering
	\includegraphics[scale=0.35]{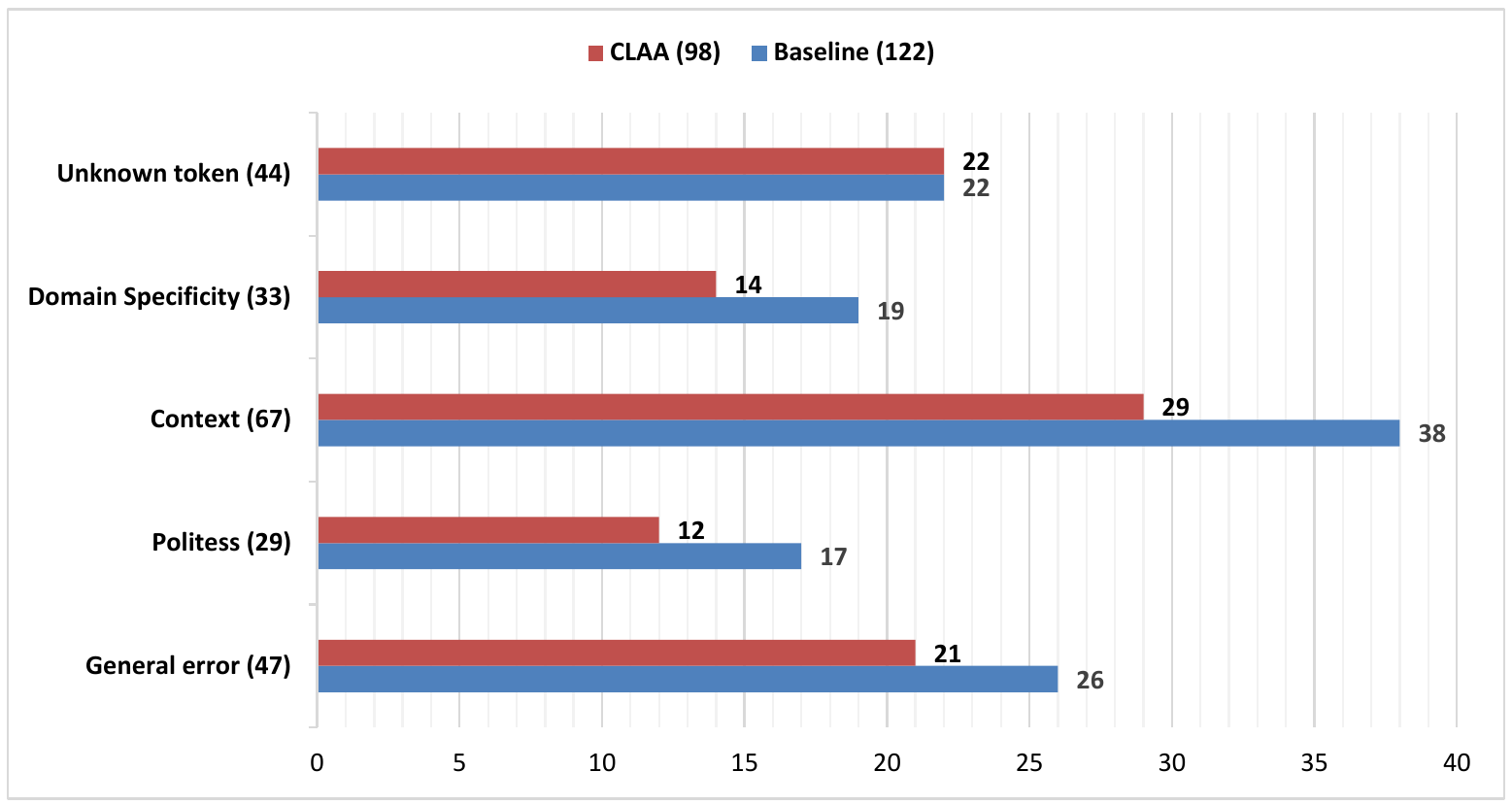}
\caption{Error categories identified from the prediction of CLAA and the baseline.}
\label{fig:comp-error}
\end{figure}

\textit{\underline{Results.}}
Figure \ref{fig:comp-error} summarizes the error categories and their distributions in our analysis. We discuss the categories below.

\noindent \textbf{General Error}. Tools made errors while processing textual contents; which led to misclassifications. These errors include failure to properly process URLs, inability to determine the presence of negation, and failure to process linguistic cues or typos. For instance, the baseline model misclassified the review \say{\textit{Not saying it isn't an issue}} because it failed to identify negation, while CLAA misclassified the review \say{\textit{I like how this works without adding more libraries to your project}} for the same reason. We also found that both models tend to misclassify reviews containing URLs, such as \say{\textit{However, I also have extensive experience in working with SWT and the URL\_http://wiki.eclipse.org/index.php/JFace [JFace- UI-toolkit], which is built on top of it}}.

\noindent \textbf{Politeness}. Tools made mistakes and categorized reviews into the wrong aspects due to the presence of polite language. For example, the baseline model misclassified reviews like \say{\textit{Thanks for the comments}} and \say{\textit{@Barak Schiller Thanks for posting link to XStream!}} because it focused on the word ``Thanks'' and ignored the reason for commenting and posting on a community. Similarly, CLAA misclassified the review \say{\textit{thanks :) jaas book is really good}} because it focused on polite markers like "thanks" and a smiley face emoticon.

\noindent \textbf{Inability to deal with context}. Tools failed to grasp the context of the text, which was essential in determining the API aspect. For instance, the baseline model incorrectly classified the sentence \say{\textit{Java Cryptography Extensions (The Practical Guide Series) by Jason Weiss"}} as it only mentions a book related to Java cryptography extensions, without providing any specific information about an API aspect. However, CLAA may have relied on keywords like "Java" and "cryptography" to predict the correct aspect (\textit{Security}) without considering the broader context. Both the baseline and CLAA misclassified the review \say{\textit{Why take the whole kitchen sink when all you need is the tap?}} as they failed to understand its context. Although the sentence uses a plumbing analogy, it does not provide any information about a specific aspect of software development or an API.

\noindent \textbf{Lack of domain knowledge}. Sometimes, tools misclassified text due to their limited understanding of the specific terminologies, jargons, and documentations related to a particular domain. For instance, in the sentence \say{\textit{Hibernate uses ANTLR for sql and hql parsing}}, the CLAA model misclassified it as it lacks the knowledge of the importance of compatibility between different software components. Similarly, in the sentence \say{\textit{CBC encryption in itself is not thread safe}}, the baseline model failed to recognize it as a security aspect related to cryptography and encryption due to the lack of domain-specific knowledge. However, CLAA model might have relied on the word ``encryption'' to make the correct prediction. In another example, \say{\textit{Look and Feel: AWT components more closely reflect the look and feel of the OS they run on}} is a sentence that talks about portability aspect. The baseline model misclassified this sentence due to its limited understanding of the challenges of software portability and the importance of designing software that can run on different platforms.

\noindent \textbf{Unknown tokens}. The tools struggled with understanding special characters like `<>', `@', and underscores in tags, class or method names. These characters were often identified as unknown tokens (<unk>). For instance, when the sentence \say{\textit{If CSS is in a <style> tag, it will be interpreted as text}} was presented to the models, both the baseline and CLAA misclassified it because they failed to recognize that `<style>' is a tag used in HTML to style documents. Transformer models depend on word patterns and frequency to grasp context, and in this case, the models might not have had enough information to differentiate between `<style>' as a tag and plain text. Another example is \say{\textit{JAXB Bindings File Sets @XmlElement type to String instead of XMLGregorianCalendar}} which was misclassified by both the models because they did not understand that `@XmlElement' is a method name. In another case, the models might have misclassified the sentence \say{\textit{Note that the example above uses a simplified way to issue calls via the \_ClientResource\_ class}} because they did not recognize that `\_ClientResource\_' is the name of a class due to the underscores in its name.

\section{Impact Analysis of CLAA (RQ2)}
We answer the following sub-RQs to assess the impact:\\
1. Does CLAA offer improvement over the baselines in terms of generalization performance (performance over unseen data)?\\
2. How useful is CLAA to support users during API selection?

\subsection{RQ2.1 Does CLAA offer improvement over the baselines in terms of generalization performance?}

\textit{\underline{Approach.}} We collected all ``json'' and ``java'' related posts from Stack Overflow and applied both CLAA and the best performing baseline to them. For this comparison, we used the RoBERTa transformer model in both the CLAA and baseline approach. Each model generated a list of sentences labeled as an aspect. We then selected a random subset of the collected sentences and manually compared the accuracy of each model on these sentences. 

\textit{\underline{Results.}}
Table \ref{tab:emp-aspect-distrib} displays the distribution of sentences among the eleven aspects labeled by both CLAA and the baseline model. The row labeled "None" indicates the number of sentences that neither model could categorize. It is evident from the table that CLAA outperformed the baseline model in most aspects, except for Bug. Specifically, in the Bug aspect, CLAA labeled 4,492 sentences, while the baseline model labeled 5,765 sentences. In some aspects such as Performance, Community, Portability, and Legal, the difference in the number of labeled sentences was relatively minor. 
It is also worth noting that CLAA categorized more sentences into multiple aspects, while the baseline model was more conservative in its labeling. Specifically, the baseline model labeled 9,196 sentences into two aspects and 180 sentences in three aspects, while CLAA labeled 10363 sentences into two aspects, 214 sentences into three aspects, and only 3 sentences into four aspects. 
\begin{table}[h]
\centering
\caption{Comparison of aspect distribution after applying pre-trained RoBERTa model in both CLAA and baseline on empirical experimental dataset consisted of 86,853 sentences. The \textit{None} row indicates the number of sentences that neither model could categorize }
\label{tab:emp-aspect-distrib}
\begin{tabular}{|c|c|c|} 
\hline
\textbf{Aspects} & \textbf{~CLAA~} & \textbf{~Baseline~}  \\ 
\hline
Performance      & 1522            & 1492                 \\ 
\hline
Usability        & 25234           & 22645                \\ 
\hline
Security         & 1407            & 1067                 \\ 
\hline
Community        & 546             & 410                  \\ 
\hline
Compatibility    & 0               & 0                    \\ 
\hline
Portability      & 336             & 274                  \\ 
\hline
Documentation    & 3515            & 3088                 \\ 
\hline
Bug              & 4492            & 5765                 \\ 
\hline
Legal            & 204             & 177                  \\ 
\hline
OnlySentiment    & 2255            & 1807                 \\ 
\hline
Others           & 47310           & 47098                \\ 
\hline
None             & 32              & 3030                 \\
\hline
\end{tabular}
\end{table}
The table also shows that the baseline model left around 3030 sentences unlabeled, while CLAA managed to label all but 32 sentences. To further investigate, we randomly selected 200 sentences, manually labeled them (the first author performed the initial annotations, which were subsequently reviewed by an external annotator with expertise in the SE domain) and calculated the accuracy of both models. CLAA achieved 92\% accuracy by correctly predicting 184 sentences, while the baseline achieved 81.5\% accuracy by correctly predicting 163 sentences. These findings provide evidence that CLAA is a more accurate and reliable model than the baseline. Despite the overall high accuracy of both models, they were unable to categorize any sentences in the compatibility aspect, indicating a limitation in their ability to categorize certain types of reviews. This could be due to the models being trained on a limited number of API reviews from Stack Overflow, which may not be representative of reviews from other sources or unseen APIs. 

\subsection{RQ2.2 How useful is CLAA to support users during API selection?}

\textit{\underline{Approach.}} We evaluated the usefulness of CLAA in API selection through a user study. 
The objective of the study was to assess if CLAA helps the users to be more accurate and confident in their API selection. After collecting the responses of the participants, we analyzed them along two dimensions: correctness and confidence. Correctness refers to the precision of the participant's selection in both settings, while confidence refers to the level of confidence they had in making their selection. Furthermore, we computed a conversion rate for each participant, which represents the ratio of participants who made an incorrect selection while using \textit{Stack Overflow} but made a correct selection when using \textit{Stack Overflow + baseline} and \textit{Stack Overflow + CLAA} separately.

\begin{table}[h]
\centering
\caption{Progression of learning from Stack Overflow to CLAA.}
\label{tab:impact}
\resizebox{\columnwidth}{!}
{\begin{tabular}{ccccc} 
\toprule
\textbf{~ ~T~ ~}   & \textbf{~ Tool~~} & \textbf{~ Correctness~~} & \textbf{~ Confidence~~} & \textbf{~ Conversion~~}  \\ 
\hline
\multirow{3}{*}{1} & SO                & 40\%                     & 3.5                     & \_                       \\
                   & ~ SO+Baseline~~   & 80\%                     & 4.4                     & 66.67\%                  \\
                   & SO+CLAA           & 100\%                    & 4.7                     & 100\%                    \\ 
\hline
\multirow{3}{*}{2} & SO                & 60\%                     & 4.2                     & \_                       \\
                   & SO+Baseline       & 90\%                     & 4.5                     & 75\%                     \\
                   & SO+CLAA           & 100\%                    & 4.8                     & 100\%                    \\
\bottomrule
\end{tabular}}
\end{table}

\textit{\underline{Results.}} All ten developers successfully accomplished their assigned tasks. In Table \ref{tab:impact}, we have included the impact of CLAA on task completion. In the case of Task 1, only 40\% of the developers who solely used \textit{Stack Overflow} were able to choose the correct API, while 80\% were successful when they used \textit{Stack Overflow + baseline}. However, when they switched to using \textit{Stack Overflow + CLAA}, all of them were able to select the right API, resulting in a 100\% conversion rate. For Task 2, 60\% of the developers using \textit{Stack Overflow} alone were able to pick the right API, while 90\% were able to do so when they used \textit{Stack Overflow + baseline}. Nonetheless, all of them were able to choose the correct API when they used \textit{Stack Overflow + CLAA}, which also resulted in a 100\% conversion rate. Additionally, the developers reported feeling more confident when utilizing CLAA with Stack Overflow to make their API selections. In the case of Task 1, their confidence level increased from 3.5 to 4.7, which is considered almost fully confident. Similarly, for Task 2, their confidence level increased from 4.2 to 4.8, indicating a high level of confidence. The follow-up survey conducted with the developers indicated that they found CLAA to be the most useful tool for API selection. For instance, P3 (i.e. participant number 3) commented that: \say{\textit{I really appreciate how CLAA helps me make a decision on which API to choose based on different aspects. It's like having a personal AI assistant for API selection!}} P7 found the visualization feature to be particularly useful in aiding decision-making, stating that \say{\textit{It makes the decision-making process much easier and quicker.}} The participants also found the usage of CLAA to be an easier and faster method for API selection. According to P10, \say{\textit{CLAA has the potential to save a lot of time and effort for developers. Instead of spending hours reading through reviews and trying to make a decision, they could use CLAA to quickly narrow down their options.}} Nonetheless, participants offered suggestions to enhance the usefulness of CLAA. For example, P1 stated that \say{\textit{I would love to see CLAA expand to cover more aspects beyond the current ones. It would make it an even more comprehensive tool for API selection.}} P8 proposed the addition of a comparison feature, which would allow developers to compare two APIs side by side based on a given aspect.

\begin{table*}[h]
\centering
\caption{LIME generated explanation for pre-trained ALBERTA model in case of both baseline (B) and CLAA (C) with highlighted features. Orange color signifies features crucial for considering a text related to an aspect, while blue color represents the opposite. The lighter the color, the less importance the feature has.}
\label{tab:lime-example}
\resizebox{\textwidth}{!}
{\begin{tabular}{cccc} 
\toprule
Aspect & Prediction & Baseline (B) & CLAA (C)  \\ 
\midrule
Usability  & \begin{tabular}[c]{@{}c@{}}Incorrect by B\\Correct by C\end{tabular}             & \adjustbox{valign=c}{\includegraphics[width=7.5cm, height=0.8cm]{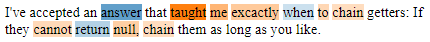}} & \adjustbox{valign=c}{\includegraphics[width=7.5cm, height=0.8cm]{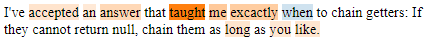}}     \\ 
\midrule
Not Usability   &   \begin{tabular}[c]{@{}c@{}}Correct by B\\Incorrect by C\end{tabular}          & \adjustbox{valign=c}{\includegraphics[width=7.5cm, height=1cm]{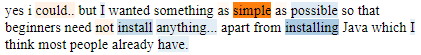}}               & \adjustbox{valign=c}{\includegraphics[width=7.5cm, height=1cm]{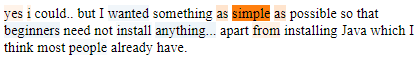}}    \\ 
\midrule
Security      & \begin{tabular}[c]{@{}c@{}}Incorrect by B\\Correct by C\end{tabular}         & \adjustbox{valign=c}{\includegraphics[width=7.5cm, height=0.8cm]{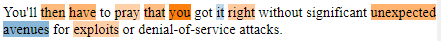}}                & \adjustbox{valign=c}{\includegraphics[width=7.5cm, height=0.8cm]{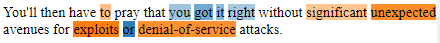}}    \\ 
\midrule
OnlySentiment            & \begin{tabular}[c]{@{}c@{}}Incorrect by B\\Correct by C\end{tabular}   & \adjustbox{valign=c}{\includegraphics[width=7.5cm, height=0.8cm]{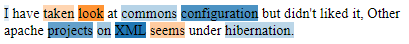}}                & \adjustbox{valign=c}{\includegraphics[width=7.5cm, height=0.8cm]{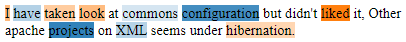}}    \\ 
\midrule
 Bug &   \begin{tabular}[c]{@{}c@{}}Incorrect by B\\Correct by C\end{tabular}           & \adjustbox{valign=c}{\includegraphics[width=7.5cm, height=1.5cm]{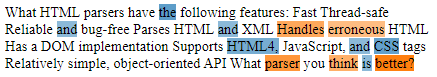}}                 & \adjustbox{valign=c}{\includegraphics[width=7.5cm, height=1.5cm]{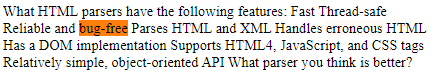}}     \\ 
\midrule
Compatibility           & \begin{tabular}[c]{@{}c@{}}Incorrect by B\\Incorrect by C\end{tabular}    & \adjustbox{valign=c}{\includegraphics[width=7.5cm, height=0.8cm]{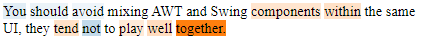}}                & \adjustbox{valign=c}{\includegraphics[width=7.5cm, height=0.8cm]{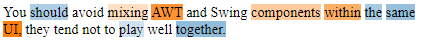}}    \\
\bottomrule
\end{tabular}}
\end{table*}

\section{Discussions}

\subsection{Feature Importance}The features emphasized by LIME when used with CLAA offers a clear understanding of the critical aspects within a review. Table \ref{tab:lime-example} presents six instances for both the baseline and CLAA, side by side, highlighting the critical features for predicting the model's outcome in a text using LIME (blue color signifies features crucial for not considering a text related to an aspect, while orange color representing the opposite). We used the ALBERT transformer model for both baseline and CLAA. Upon analyzing the predictions and the highlighted features by LIME for both models, it appears that CLAA outperformed the baseline model by accurately categorizing the aspects. The features identified by LIME using CLAA provide a precise understanding of the essential aspects of the text, whereas the features highlighted by the baseline model were not as relevant to the core aspects of the text. Due to space constraints, we limit our analysis of this claim to only the first and last examples listed in Table \ref{tab:lime-example}.
For the first example, the baseline model misjudged the review as not related to usability. The features highlighted by LIME using the baseline model for usability aspect included "taught", "me", "exactly", "to", "chain", "cannot" and "null" which are all related to the concept of "chaining". However, the model seems to have missed the crucial aspect of the discussion which is about when to chain getters. In contrast, CLAA made the correct prediction. The features highlighted by LIME using CLAA in case of usability included "accepted", "an", "answer", "taught", "me", "exactly", "long", "you" and "like" which are all related to the discussion of when to chain getters. Thus, CLAA has accurately captured the key aspects of the text.
In the second example, the baseline model correctly predicted that the text was not related to usability, while CLAA made an incorrect prediction by identifying the text as related to usability. The features highlighted by LIME using the baseline model for considering the text as not related to usability included "I", "have", "anything", "possible", "install" and "installing". These words suggested that the text was discussing installation, which is not directly related to usability. Meanwhile, the features identified by LIME using CLAA for considering the text as not related to usability included "anything", "beginners", "install", "wanted" and "people". These words also provide no clear indication of the idea related to usability. However, CLAA seems to have captured the idea that the user "wanted something as simple as possible" which might have been the reason behind the incorrect prediction.
In case of third example, the original aspect was security, and although the baseline model incorrectly predicted the text as not related to security aspect, CLAA made the correct prediction. The features highlighted by LIME for not considering the text related to security aspect using both models were "it", "avenues" and "got" which do not directly relate to the concept of security. However, CLAA identified "or" as a feature for not considering as a security aspect, which could imply that it recognized the importance of considering other possibilities or alternatives. In contrast, the features identified using the baseline model for security aspect were "you", "have", "then", "that", "right", "unexpected" and "pray" which are less relevant to the concept of security to make the correct prediction. CLAA identified "exploits", "unexpected", "denial-of-service", "to" and "significant" as features that are more directly related to the aspect of security.
Similarly, in the fourth example, the baseline model seems to focus more on technical terms such as "commons configuration", "XML" and "hibernation" which could lead to a classification of the text as not sentiment-related. On the other hand, CLAA includes more sentiment-related features such as "liked" and "taken look" that indicate a positive or negative sentiment towards the topic. In this case, the correct classification would be OnlySentiment aspect, which is correctly predicted by CLAA.
When comparing the two models in the fifth example, it appears that they have differing views on how to classify the text. The original aspect was related to bug, and while the baseline model inaccurately predicted that the text was not related to bug, CLAA correctly classified it as bug. Using the baseline model LIME highlighted "the", "and", "html4", "css" and "is" as features that do not directly relate to the concept of bugs, whereas CLAA did not identify any such features. As for the LIME features related to bug aspect, the baseline model focused on "handles", "parser", "think", "better" and "erroneous" which suggest a focus on specific language and phrasing in the text. In contrast, CLAA's features for bug aspect included "bug-free" and "erroneous" indicating a greater emphasis on specific requirements and features mentioned in the text.
Despite both the baseline model and CLAA making an incorrect prediction for the sixth example, the features extracted by LIME using both models suggest that they have recognized some significant words and expressions related to the compatibility aspect. The baseline model has identified the words "together", "well", "within", "tend" and "components" as important for compatibility aspect, while CLAA has identified "UI", "AWT", "mixing", "within" and "components". It is possible that the models might have been confused by the negation ("avoid mixing") in the sentence, which could have led them to assign a lower probability to the sentence being related to compatibility. Additionally, the sentence itself is relatively short and does not provide a lot of contextual information, which could make it difficult for the models to accurately classify the sentence.

\subsection{Threats to Validity}
\textbf{Internal validity threats} relate to the limitations in study design and implementation that may impact experimental results. To reduce these threats, we reused an existing replication package for aspect based API review classification provided by \cite{yang2022aspect} and performed necessary modification to implement CLAA. We used the contrastive loss functions provided in \cite{Musgrave2020PyTorchML}. Potential threats introduced due to the usage of LIME include sensitivity to perturbation strategy, inadequate sample size for generating explanations, noise and randomness issues, biases inherited from the original model etc. To mitigate these threats, for each explanation generation, we have have used a sample size of 100 (perturbed samples), to reduce noise and randomness we have used a seed value of 42. To validate the explanations, quantitative and qualitative analysis is need which we leave open for future work.

\textbf{External validity threats} refers to the generalizability of the results. For API aspect detection using CLAA, the results may only apply to the specific dataset used and the domain of Stack Overflow but may not be applicable to other datasets or domains. 

\textbf{Construct validity threats} refer to concerns regarding the relationship between theory and observations. Such threats could arise due to errors in measurement. We used the same evaluation metrics including precision, recall, f1 score, MCC, and AUC that were used in \cite{yang2022aspect}. 
To gauge the performance of the participants, we determined the percentage of correct answers, which may be influenced by external factors like tiredness, time management etc.

\section{Related Work}
Related work can broadly be divided into two types: studies that aimed to understand the nature and types of software and API aspects and tools that are developed to detect and analyze the aspects.

\textbf{Studies.}
Barua et al. \cite{barua2014developers} analyzed Stack Overflow and found that discussions were mainly about programming languages, tools, and frameworks, and their popularity correlated with the corresponding technologies. Uddin et al. \cite{uddin_understand} investigated how and why developers seek and analyze API-related opinions on Stack Overflow. 
Uddin and Khomh \cite{uddin2019automatic} proposed a technique for automatically mining opinions expressed about APIs in Stack Overflow; 
Lin et al. \cite{lin2019pattern} proposed a pattern-based mining technique to extract opinions from Q\&A websites, finding that developers frequently express opinions about API usability and functionality. Uddin and Khomh proposed techniques to summarize API reviews \cite{uddin2017automatic} and mine API aspects from reviews \cite{uddin2017mining}, finding that developers often express opinions about aspects like ease of use, documentation, and performance. Zhang and Hou \cite{zhang2013extracting} proposed a technique to extract problematic API features from forum discussions and found that developers often find issues related to documentation, compatibility, and complexity of APIs. Ahasanuzzaman et al. \cite{ahasanuzzaman2018classifying} aimed at detecting posts on Stack Overflow that are related to API issues. In their subsequent study, Ahasanuzzaman et al. proposed a supervised learning method named CAPS \cite{ahasanuzzaman2020caps}, which employed five distinct dimensions and a conditional random field (CRF) technique. 

\textbf{Tools.} Several recent studies have utilized natural language processing to enhance API documentation quality and classify API reviews based on their aspects. Opiner \cite{uddin2017opiner} extracts and summarizes relevant opinions about APIs from user comments on various platforms, while Treude and Robillard \cite{treude2016augmenting} automatically detect API-related sentences from Stack Overflow discussions to augment API documentation. Uddin and Khomh employ ML classifiers to identify API aspects \cite{uddin2017opiner}, and Yang et al. \cite{yang2022aspect}, Nibir et al. \cite{nibiraspect} used pre-trained transformer models for aspect-based API review classification. In contrast, CLAA uses contrastive learning prior to fine-tuning transformer models as classifiers. In the software engineering field, contrastive learning is gaining popularity, and it has been utilized in recent works such as SynCoBERT \cite{wang2021syncobert} for multi-modal code representation, ContraCode \cite{jain-etal-2021-contrastive}, Heloc \cite{wang2022heloc} for code representation learning, CODE-MVP \cite{wang-etal-2022-code} for code representation from multiple views, Clear \cite{wei2022clear} for API recommendation, varclr \cite{chen2022varclr} for variable semantic representation, contrastive learning for code clone detection \cite{khajezade2022evaluating}, bug priority inference \cite{wang2022clebpi}, multi-modal code review \cite{wu2022contrastive}, code retrieval and summarization \cite{bui2021self}. 

\section{Conclusion}
In this paper, we propose CLAA, a tool that helps with API selection and aspect-wise online reviews aggregation. It uses a two stage training: first it uses supervised contrastive training objective to fine-tune seven pre-trained transformer models to learn aspect-wise semantic representations of API review instances, which are then used to act as sequence classifiers for API aspect detection tasks. CLAA uses LIME to explain why the classifier makes certain predictions. Experimental results show that CLAA significantly outperforms the state-of-the-art baselines in terms of F1 score, MCC score, and AUC score. RoBERTa performed the best in CLAA among the seven transformer models. However, the results also showed that models with large structures may not always perform well in categorizing API reviews. There is still room for further research in applying contrastive learning to other pre-trained language models, and the dataset used in the study is imbalanced, which could affect the results. CLAA was found to be more accurate than the baseline model in online actual and developer studies. 


\bibliographystyle{IEEEtran}
\bibliography{software}


\end{document}